\documentclass[
aps, 
prl,   
reprint,
groupedaddress,
superscriptaddress,
longbibliography
]{revtex4-2}

\usepackage{indentfirst}
\usepackage{graphicx}
\usepackage{dutchcal}
\usepackage{bm}
\usepackage{amsmath,amssymb}
\begin{document}

	\title{High-Capacity and Real-Time Acoustic Communication by Multiplexing Velocity}
	\author{Lei Liu}
	\affiliation{National Laboratory of Solid State Microstructures and Department of Materials Science and Engineering, Nanjing University, Nanjing 210093, China}
	
	\author{Xiujuan Zhang\thanks{Corresponding author}}
	\email[]{xiujuanzhang@nju.edu.cn}
	\affiliation{National Laboratory of Solid State Microstructures and Department of Materials Science and Engineering, Nanjing University, Nanjing 210093, China}
	
	\author{Ming-Hui Lu}
	\affiliation{National Laboratory of Solid State Microstructures and Department of Materials Science and Engineering, Nanjing University, Nanjing 210093, China}
	\affiliation{Jiangsu Key Laboratory of Artificial Functional Materials, Nanjing 210093, China}
	\affiliation{Collaborative Innovation Center of Advanced Microstructures, Nanjing University, Nanjing 210093, China}
	
	\author{Yan-Feng Chen}
	\affiliation{National Laboratory of Solid State Microstructures and Department of Materials Science and Engineering, Nanjing University, Nanjing 210093, China}
	\affiliation{Collaborative Innovation Center of Advanced Microstructures, Nanjing University, Nanjing 210093, China}
	
	\begin{abstract}
		Acoustic communication is indispensable for underwater networks, deep ocean exploration, and biological monitoring, environments where electromagnetic waves become impractical. However, unlike the latter, whose vector polarization naturally supports multiple information channels, acoustic waves are longitudinal and have traditionally relied almost exclusively on a single scalar pressure channel, posing a fundamental limit on their data-carrying capacity. Here, we theoretically and experimentally demonstrate that the vector velocity of acoustic waves can serve as a polarization-like physical degree of freedom. Using its three components as mutually independent communication channels and demodulating them with a single vector sensor, we achieve reliable, high-capacity, and real-time information transmission. Multiplexing velocity adds a new dimension to acoustic communication. When combined with other physical degrees of freedom (frequency, phase, etc.), this approach can significantly enhance the information capacity, opening new avenues for next-generation acoustic technology.
	\end{abstract}

	%\keywords{}
	
	\maketitle
	\textit{Introduction}---As human technology expands into challenging environments, such as deep oceans, biological tissues, and other opaque media, optical links become impractical due to strong absorption and severe scattering of electromagnetic waves\cite{hale1973optical}. Particularly in underwater communication, mechanical acoustic waves remain the only option for transmitting information over considerable distance\cite{kilfoyle2002state,stojanovic2002recent}. However, the low operational frequency and slow propagation speed of acoustic waves severely restrict their communication capacity\cite{akyildiz2005underwater,stojanovic2009underwater,li2025underwater}. Although multiplexing techniques with time, amplitude, frequency, and phase have significantly boosted data transmission rates\cite{1090990,stojanovic1994phase,lidstrom2023evaluation}, communication capacity of acoustic waves is still highly limited by their longitudinal nature that imprints a scalar pressure field.
	
	Recent research has made significant efforts to develop orbital angular momentum (OAM) as a new degree of freedom for acoustic multiplexing communication\cite{marston2010modulated,shi2017high,jiang2018twisted,li2020principle,jia2022orbital,wu2022metamaterial,zhang2023spatiotemporal}. Due to the orthogonality of OAM modes with different topological charges, independent communication channels can coexist without intermodal crosstalk. However, the deployment of OAM-based acoustic communication remains limited by two major challenges. First, OAM modulation and demodulation typically require bulky and complex transducer arrays or artificial structures\cite{shi2017high,jiang2018twisted,li2020principle,jia2022orbital,wu2022metamaterial}, which hinder miniaturization and system integration. Second, vortex beams carrying OAM are highly susceptible to diffraction, leading to rapid spatial divergence and significant loss of transmission fidelity over long distance\cite{paterson2005atmospheric,xie2015performance}.

	Beyond the scalar pressure and OAM degrees of freedom, the vector acoustic velocity has recently emerged as a polarization-like resource\cite{muelas2022observation}. Although acoustic waves are longitudinal and curl-free, acoustic spin, arising from the local rotation of the velocity field, has been theoretically identified and experimentally observed in interference patterns, evanescent fields, and artificial structures\cite{long2018intrinsic,shi2019observation,bliokh2019spin,bliokh2021spatiotemporal,wang2021spin,yang2023hybrid,liu2025cyclic,tan2025revealing,liu2025skyrmion}. Unlike scalar pressure, vector velocity intrinsically carries directional information, offering three additional degrees of freedom in three-dimensional space. This raises an intriguing question: can these velocity-based degrees of freedom be harnessed to increase acoustic communication capacity?
	
	\begin{figure}[htbp]
		\centering
		\includegraphics[width=1\linewidth]{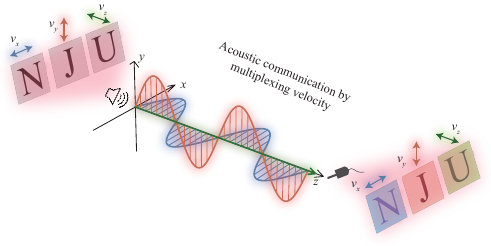}
		\caption{\label{fig1}
			Schematic of multi-channel acoustic communication by multiplexing velocity. Three images (“N,” “J,” and “U”) are individually encoded onto the three orthogonal velocity channels $v_x$, $v_y$, and $v_z$. After parallel transmission, the signals are simultaneously decoded at a single-point receiver, fully reconstructing the original images.	
		}
	\end{figure} 
	Here, we demonstrate that the answer is affirmative. By leveraging the three orthogonal components ($v_x$, $v_y$, and $v_z$) of acoustic velocity as independent communication channels---analogous to polarization multiplexing in optics---we achieve reliable, high-capacity, and real-time information transmission with minimal hardware. The method rests on a direct mapping between pressure multipole modes and individual velocity components, allowing each velocity channel to be modulated independently. At the receiver, a three-dimensional velocity sensor enables simultaneous demodulation of all three components at one spatial point. Experimentally, we demonstrate parallel transmission of three distinct images through the three velocity channels (as schematically shown in Fig.~\ref{fig1}), effectively tripling the communication capacity compared to a conventional scalar pressure channel. We further assess the communication performance in the presence of environmental noise and channel imperfections, confirming excellent accuracy and stability. 
             
     We introduce vector velocity as an additional, independent multiplexing dimension in acoustic communication, which not only overcomes the fundamental capacity limit inherent to scalar fields but also delivers a compact and efficient demodulation strategy. Fully compatible with existing modulation schemes, such as amplitude, phase, frequency, etc., this scheme is immediately integrable with current acoustic systems and provides a practical route toward higher-capacity underwater and biomedical acoustic links.
	
	\begin{figure}[htbp]
	\centering
	\includegraphics[width=1\linewidth]{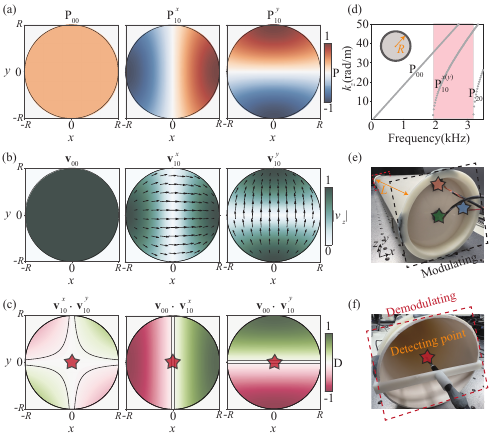}
	\caption{\label{fig2}
		(a) $x$–$y$ cross-sections of the pressure field distributions for the three lowest-order modes of a cylindrical waveguide, i.e., one monopole and two dipole modes. 
		(b) Corresponding velocity field distributions. The arrows indicate $v_x$ and $v_y$.
		(c) Normalized inner product $D$. The black contour encloses $\left | D \right | <0.05$, confirming near-perfect orthogonality. 
		(d) Dispersion relation of a rigid-walled air cylindrical waveguide. Red-shaded region indicates the frequency window where only the monopole and dipole modes propagate.
		(e) Modulation scheme: three acoustic transducers at $(x,y)=(0.85R,0)$ (blue star), $(0,0.85R)$ (yellow star), and $(0,0)$ (green star) on the input port independently modulate $v_x$, $v_y$, and $v_z$ channels, respectively.
		(f) Demodulation scheme: a three-dimensional velocity sensor positioned at the waveguide center (red star) on the receiving side simultaneously decodes all three velocity channels.
	}
    \end{figure}
	\textit{Modulation–demodulation for the three velocity channels}---In any communication system, efficient modulation and demodulation determine both data-rate and fidelity. For acoustic communication based on multiplexing velocity, accurate control and independent readout of individual velocity components are essential. We start with the linear Euler equation for acoustic wave in a homogeneous dense medium (liquid or gas)\cite{pierce2019acoustics}
	\begin{equation}
		\rho \frac{\partial \mathbf{v} }{\partial t} =-\nabla P,
		\label{eq1}
	\end{equation}
	where $P$ is the pressure field, $\mathbf{v} (v_x, v_y, v_z)$ the velocity field, and $\rho$ the mass density. For a monochromatic wave of angular frequency $\omega$, assuming a time-harmonic dependence \(e^{-i\omega t}\), Eq.~(\ref{eq1}) is reduced to
	\begin{equation}
		\mathbf{v} =-\frac{i}{\rho \omega } \nabla P.
		\label{eq2}
	\end{equation}
    Equation~(\ref{eq2}) reveals that the three velocity components ($v_x,v_y,v_z$) are generally coupled through the spatial derivatives of the pressure field and thus cannot be manipulated separately within a single pressure mode. 
    
    To achieve independent control of the velocity components, we employ a cylindrical waveguide which supports orthogonal multipole modes, each coupling predominantly to one velocity component. In polar coordinates $(r,\varphi)$, the eigenmodes can be expressed as\cite{pierce2019acoustics,hartig1938transverse,kinsler2000fundamentals}
    \begin{equation}
    	P_{mn} = A_{mn} J_m(k_{mn} r) \begin{Bmatrix}
    		\cos m\varphi  \\\sin m\varphi \end{Bmatrix} e^{i (k_z z - \omega t) },
    	\label{eq3}
    \end{equation}
    where $A_{mn}$ denotes the modal amplitude. $J_m$ is the first-kind Bessel function of $m$th order. $k_{mn}$ is the radial wave vector component, which can be determined by the Neumann boundary condition $[\partial J_m(k_{mn} r)/{\partial r}]_{r=R}=0$, with $n$ denoting the $n$th root. It obeys the dispersion relation $k_{mn}^2+k_{z}^2=\omega ^2/c^2$, where $k_{z}$ is the propagating wave vector component and $c$ is the acoustic speed in the medium. For the three lowest‐order modes, we have  
    \begin{subequations}\label{eq4}  
    	\begin{align}
    		P_{00}     &= A_{00}\,e^{i(k_z z - \omega t)}, \label{eq4a}\\
    		P_{10}^{x} &= A_{10}^{x}\,J_1(k_{10} r)\cos \varphi\, e^{i(k_z z - \omega t)}, \label{eq4b}\\
    		P_{10}^{y} &= A_{10}^{y}\,J_1(k_{10} r)\sin \varphi\, e^{i(k_z z - \omega t)}. \label{eq4c}
    	\end{align}
    \end{subequations}
    With the modal amplitudes normalized to unity, Fig.~\ref{fig2}(a) presents the corresponding $x$–$y$ cross-sections. $P_{00}$ corresponds to a monopole plane-wave-like mode, whereas $P_{10}^{x}$ and $P_{10}^{y}$ form a degenerate dipole pair with orthogonal dipole moments. The orientation of each dipole moment depends on the excitation protocol. Without loss of generality, we assign them along the $x$- and $y$-axes.
    
    \begin{figure*}[htbp]
    	\centering
    	\includegraphics[width=1\linewidth]{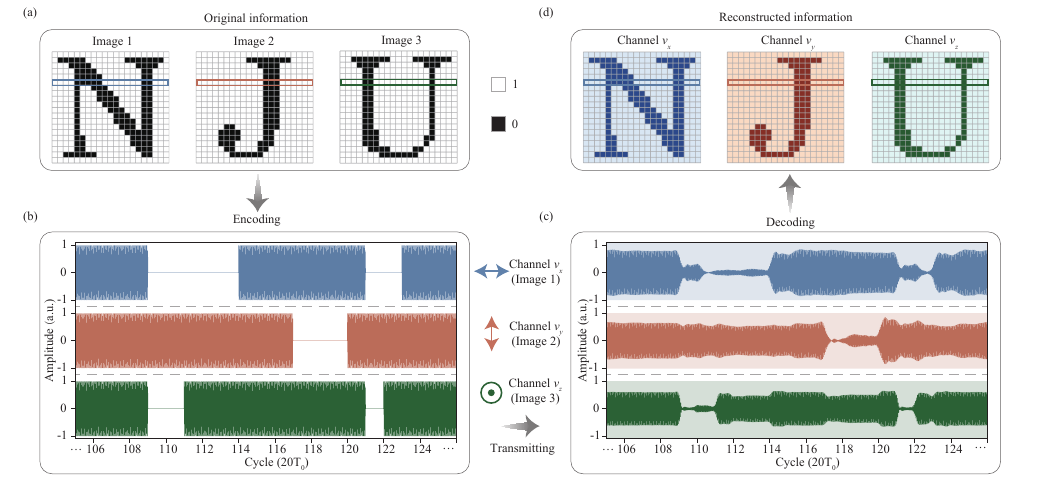}
    	\caption{\label{fig3}
    		(a) Binary images of the letters “N,” “J,” and “U,” each discretized into $21 \times 20=420$ pixels, with white and black pixels representing logic “1” and “0,” respectively.
    		(b) Input signal segments for the sixth rows of the images in (a), which are successively encoded into the $v_x$(blue), $v_y$(yellow), and $v_z$(green) channels.
    		(c) Time-domain output signals received at the far end by a one-shot, single-point measurement using a velocity sensor.
    		(d) Reconstructed images from the received data streams.
    	}
    \end{figure*}
    The velocity distributions are obtained from Eq.~(\ref{eq2}) and shown in Fig.~\ref{fig2}(b). The color map represents $|v_z|$, while the arrows denote $v_x$ and $v_y$.  It is evident that the velocity fields corresponding to $P_{00}$, $P_{10}^{x}$, and $P_{10}^{y}$ are dominated by $v_z$, $v_x$, and $v_y$, respectively. Explicitly, at the waveguide center ($r=0$), these three modes are perfectly orthogonal to each other, yielding independent velocity components
    \begin{subequations}\label{eq5}  
    	\begin{align}
    		\mathbf{v}_{00}&=
    		\begin{pmatrix}
    			0, & 0, & \frac{k_z}{\omega\rho}\,A_{00}\,e^{i(k_z z - \omega t)}
    		\end{pmatrix}, \label{eq5a}\\
    		\mathbf{v}_{10}^{x}&=
    		\begin{pmatrix}
    			\frac{k_{10}}{2\omega\rho}\,A_{10}^{x}\,e^{i(k_z z - \omega t + \frac{\pi }{2})}, & 0, & 0
    		\end{pmatrix}, \label{eq5b}\\
    		\mathbf{v}_{10}^{y}&=
    		\begin{pmatrix}
    			0, & \frac{k_{10}}{2\omega\rho}\,A_{10}^{y}\,e^{i(k_z z - \omega t + \frac{\pi }{2})}, & 0
    		\end{pmatrix}. \label{eq5c}
    	\end{align}
    \end{subequations}
    To quantify their orthogonality, we calculate the normalized inner product $D = \frac{\mathbf{v}_j \cdot \mathbf{v}_k}{\left | \mathbf{v}_j \right | \left | \mathbf{v}_k \right |}$ (with $j$ and $k$ the modal order), as shown in Fig.~\ref{fig2}(c). The three velocity fields are indeed mutually perpendicular at the waveguide center where $D = 0$ (as marked by the red stars). Such perfect orthogonality enables independent excitation and fully decoupled modulation of all three velocity components, which are fundamental to parallel transmission and reliable readout in communications.
    
    The foregoing analysis holds for any homogeneous medium, including both air and water. As a proof of concept, we validate the proposed scheme by employing an air cylindrical waveguide of radius $R=5.25\,\mathrm{cm}$ [see the inset of Fig.~\ref{fig2}(d)]. Its dispersion relation [Fig.~\ref{fig2}(d)] reveals one monopole and two dipole modes propagating within the frequency band of 1.9–3.2 kHz (highlighted by the red-shaded region). For signal modulation, three acoustic transducers [donated by the stars in Fig.~\ref{fig2}(e)] are carefully positioned according to the distinct field profiles of the three modes. Specifically, they are placed at $(x,y)=(0.85R,0)$ (blue star), $(0,0.85R)$ (yellow star), and $(0,0)$ (green star) to independently modulate $v_x$, $v_y$, and $v_z$, respectively (details in Section \uppercase\expandafter{\romannumeral1} of Supplementary Material\cite{supp}). For demodulation, we use a three-dimensional velocity sensor to receive the transmitted signals at the waveguide center on the far end, as shown in Fig.~\ref{fig2}(f). Such sensor is capable of simultaneously measuring all three velocity components (see Section \uppercase\expandafter{\romannumeral2} of Supplementary Material\cite{supp}), enabling a single-point demodulation.

	\textit{Real-time communication using velocity multiplexing}---Building on the modulation–demodulation scheme, we experimentally demonstrate real-time image transmission using binary amplitude-shift keying (2ASK) carried on the three velocity channels.
	
    Figure~\ref{fig3}(a) shows the binary images of the letters “N,” “J,” and “U,” each discretized onto a $21 \times 20$ pixel grid (420 bits). Using logic “1” and “0” (represented by the white and black pixels, respectively), three parallel bitstreams are generated. These streams are subsequently encoded onto 2.5 kHz acoustic carrier waves (with period $T_0=0.4\,\mathrm{ms}$), for which the waveguide supports only the three lowest-order modes. Each stream is pulse-modulated, with one bit conveyed by a pulse cycle lasting $20\,T_0=8\,\mathrm{ms}$. Cycles 106–125, corresponding to the sixth row of each image, are displayed in Fig.~\ref{fig3}(b). The three modulated signals are simultaneously fed to transducers located at $(x,y)=(0.85R,0)$ (blue), $(0,0.85R)$ (yellow), and $(0,0)$ (green), thereby encoding the image information onto the three velocity channels. This procedure yields a data rate of $125\,\mathrm{bit\,s^{-1}}$ per channel and an aggregated rate of $375\,\mathrm{bit\,s^{-1}}$, tripling the capacity attainable using a conventional pressure channel. The straightforward 2ASK modulation readily confirms the capability of velocity-multiplexing to enhance the data capacity. Further integration with advanced techniques such as multiple-input-multiple-output (MIMO) processing or high-order shift-keying could potentially surpass current tens of $\mathrm{kbit\,s^{-1}}$ limits in underwater acoustic communications\cite{li2009mimo,tao2010robust,kida2023experiments}.
	
	After propagating $L=1.1\, \mathrm{m}$ ($\approx 8\lambda$), the signals are demodulated. The received time-domain data by our velocity sensor is shown in Fig.~\ref{fig3}(c) in segments, faithfully reproducing the input signals of Fig.~\ref{fig3}(b). Because the velocity sensor is capably of one-shot, single-point detection of all three velocity components, decoding can be executed immediately upon reception, enabling a real-time demodulation without the latency from time-consumed scanning and postprocessing. The reconstructed images are depicted in Fig.~\ref{fig3}(d). The corresponding bit error rates (BERs), defined by the ratio of erroneous bits to total transmitted bits, are measured as 0, 0 and 0.002 for the $v_x$, $v_y$, and $v_z$ channels, respectively, confirming high-fidelity image transmission via velocity multiplexing. 
	
	\begin{figure}[htbp]
		\centering
		\includegraphics[width=1\linewidth]{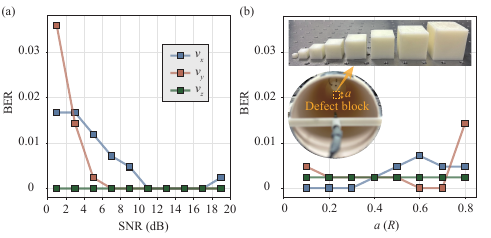}
		\caption{\label{fig4}
			 Measured BER for the $v_x$, $v_y$, and $v_z$ channels versus (a) SNR (indicating the ambient noise level) and (b) size of an obstacle inserted midway along the waveguide (inset), demonstrating high-fidelity performance of the velocity-multiplexed acoustic communication.
		}
	\end{figure}
	\textit{Robustness analysis}---To quantitatively evaluate the fidelity of the velocity–multiplexed acoustic communication, we measure BERs under controlled ambient noise and channel defects. 
	
	First, we add white noise of varying levels to the input signals and measure its influence on the BER. Noise intensity is quantified by the signal-to-noise ratio (SNR), defined as $10\log_{10}(W_S/W_N)$, where $W_S$ and $W_N$ denote the signal and noise power, respectively. For each SNR level, we transmit the three images in the same manner as in Fig.~\ref{fig3} and record the BER of the reconstructed images, as shown in Fig.~\ref{fig4}(a). Across all velocity channels, the BER remains below 0.035 and decreases with increasing SNR, effectively dropping to zero when the SNR exceeds $11\;\mathrm{dB}$. It is worth noting that the $v_z$ channel is more resilient than $v_x$ and $v_y$ because it is carried by the fundamental plane-wave mode, whereas $v_x$ and $v_y$ rely on higher-order dipole modes which are intrinsically more sensitive to disturbances.
	
	Next, we assess the influence of channel defects by inserting a rigid square obstacle of side length $a$ at the waveguide midpoint, as shown by the inset of Fig.~\ref{fig4}(b). Repeating the same transmission procedure while progressively enlarging $a$, we obtain the BER curves as presented in Fig.~\ref{fig4}(b). All channels maintain BER below $10^{-2}$ even when $a$ reaches $0.8R$, confirming reliable operation despite significant channel defects. Consistent with the noise study, the $v_z$ channel again exhibits the highest stability, followed by $v_x$ and $v_y$.
	
	\textit{Conclusions}---We have established velocity multiplexing as a novel strategy for reliable, high-capacity and real-time acoustic communication. By treating the three orthogonal velocity components as independent data channels, we transmit three binary images  in parallel and decode them with a one-shot, single-point measurement, achieving BERs below 0.002 and maintaining high fidelity under strong ambient noise and sizable channel defects. 
	
	Framing velocity as a polarization-like degree of freedom expands the physical channels available to acoustic communication. In combination with conventional dimensions (such as phase, frequency, amplitude, etc.) and advanced digital modulation technologies (e.g., high-order shift keying), velocity multiplexing can substantially boost the data capacity. Crucially, as a strictly local quantity, velocity enables demodulation of all data channels at one spatial point for one shot, facilitating compact transceiver designs that avoid bulky sensor arrays or complex metamaterials. 
    
    Our approach can be readily extended to underwater settings. Naturally-occurring acoustic ducts formed by depth‐dependent seawater density stratification can support the required guided modes and enable velocity‐based communication in the ocean\cite{stojanovic2009underwater,tadayon2019iterative}. More broadly, the demonstrated control of vector velocity in acoustics opens a pathway for adopting polarization-based techniques—such as secure encryption, imaging, and sensing—within the acoustic domain.
    
    \par\vspace{1\baselineskip} 
    \raggedbottom 
    \begin{acknowledgments}
		\textit{Acknowledgments}---This work is supported by the Key R\&D Program of Jiangsu Province (Grant No. BK20232015), the National Natural Science Foundation of China (Grant No. 12222407), and the National Key R\&D Program of China (Grants No. 2023YFA1407700 and No. 2023YFA1406904). We thanks Xiang-Yuan Xu and Yuan Sun for helpful discussions.
	\end{acknowledgments}

\end{document}